# Utilizing the Active and Collaborative Learning Model in the Introductory Physics Course

Nguyen Hoai Nam[1]

[1] Faculty of Technology Education, Hanoi National University of Education, Hanoi, Vietnam

Correspondence: Nguyen Hoai Nam, 136 Xuanthuy Caugiay Hanoi Vietnam. Tel: 84-912-172-474. E-mail: namnh@hnue.edu.vn

**Abstract**

Model of active and collaborative learning (ACLM) applied in training specific subject makes clear advantage due to the goals of knowledge, skills that students got to develop successful future job. The author exploits the learning management system (LMS) of Hanoi National University of Education (HNUE) to establish a learning environment in the Introductory Physics course Part 2, which supports to the blended learning, for the 1st year student. The method of pattern languages is taken to redesign learning content in the framework of ACLM, which focused on analysing real-life application and doing exercises, with the support of information and communication technology (ICT). Both formative and summative assessments were used to evaluate competency of students.

An online survey in middle course was used to collect 82 responses of students in closed with 63 questions and 6 open forms which indicated not only soft skills of students but knowledge of contents and technical enhanced. The result suggested that students should be attached to weekly tasks for doing work properly and practising to overcome negative attitudes to collaboration. Art performance in the welcome or the rest of the lecture was a solution to increase students' learning interest.

The conclusion recommended that deploying ICT in combination of social network and LMS with blended learning mode will increase the efficiency of ACLM so it need to be studied and supported.

**Keywords:** pedagogy, active and collaborative learning, technology education, introductory physics, training credit, learning environment

## 1. Introduction

Competency-based learner training is most concerned by Vietnam Education to meet the requirement of innovation. For the success, factors of training hold a great of attention which need changing, such as: content, curriculum, methodology, technology and so on.

Hanoi National University of Education (HNUE) is a key university of training teacher, therefore it is asked to study solutions to enhance quality of future instructor to meet education reform requirement.

For several years, the information and communication technology (ICT) facilities of HNUE have been invested which brings about a great opportunity for changing teaching methodology whereas a modern network mostly supported for intranet/internet exploitation and related services. Deploy learning management system of HNUE in training will enhance the efficiency of the network and learning.

## 2. Methodology

### *2.1 TPACK & TLCK Framework*

Since ICT applied in teaching and learning, many advantages have been confirmed (Smeets & Mooij, 2001; Ruthven et al., 2004; Smeets, 2005). However, lack of ICT prepared knowledge made difficulties for teacher to design an effective lecture (Brush & Saye, 2009; Kramarski & Michalsky, 2010). To help teachers, a theoretical framework was developed (Koehler & Mishra, 2005; Mishra & Koehler, 2006) and contributed by many scholars. It was first called by TPCK (technological pedagogical content knowledge) then changed to TPACK for the ease of pronunciation (Thompson & Mishra, 2007-2008) which build on Shulman's descriptions of effective teaching with technology (Shulman, 1986, 1987). According to the TPACK's framework, there are three major forms of knowledge of teacher required as Content (CK), Pedagogy (PK), and Technology (TK) demonstrated as circles. The four interrelated kinds of knowledge also created by the overlap of the circles as Pedagogical content knowledge (PCK), Technological content knowledge (TCK), Technological pedagogical knowledge (TPK) and Technological pedagogical content knowledge (TPCK) (Mishra and Koehler, 2009). The definition of these terms can be found in the works of Koehler and Mishra (2005; 2006; 2009). Many researches pointed out the success of technology implementation in teaching (Brouwer et al., 2009; De Laat, et al. 2007; Garrison & Vaughan, 2008; Giesbers et al., 2013; Rienties & Townsend, 2012; Rienties et al., 2013). Figure 1 below shows diagramatic depiction of TPACK framework.

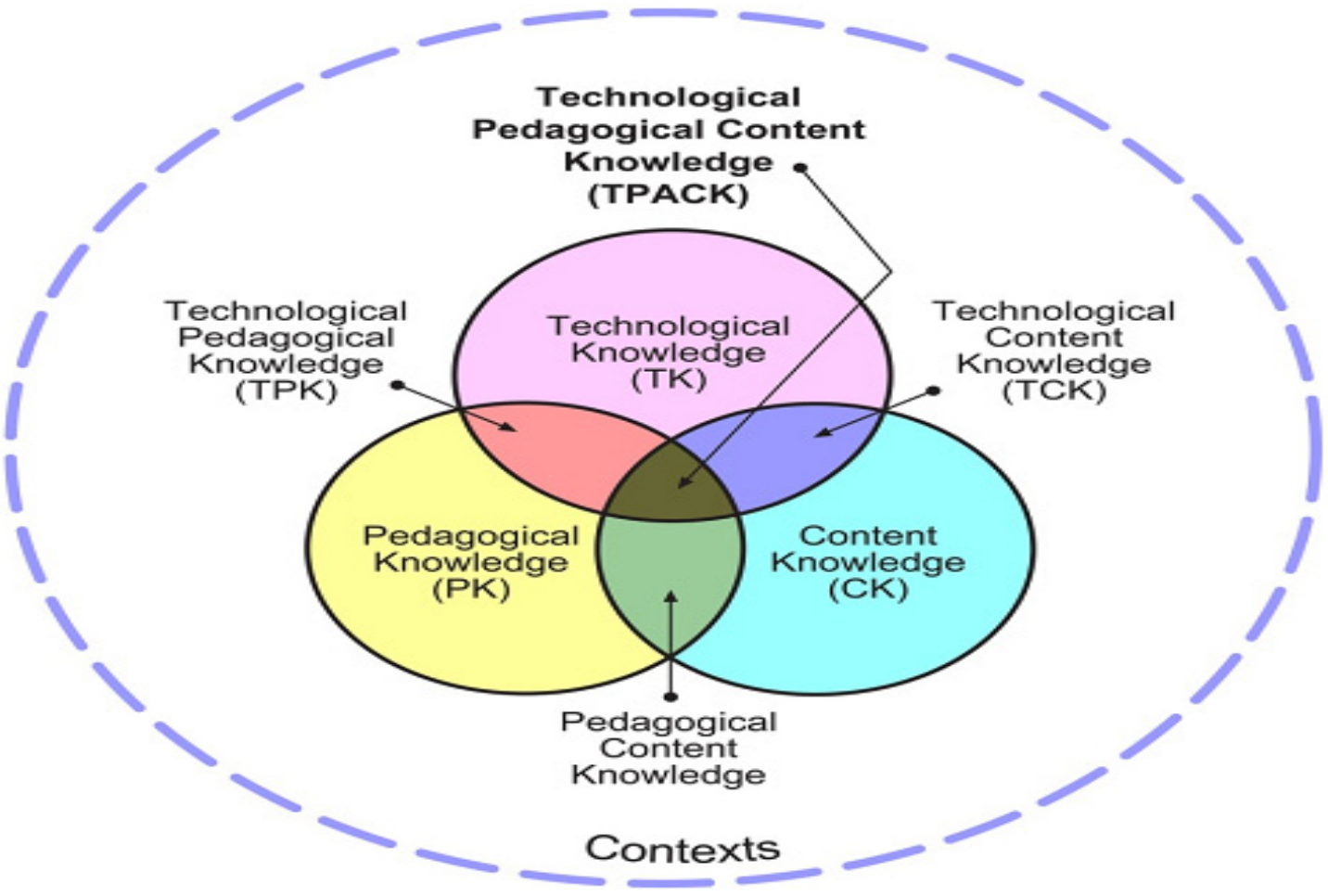

Figure 1. TPACK framework. The three circles: Content, Pedagogy, and Technology. Overlap to lead to four more kinds of interrelated knowledge. (Figure 1. In Koehler & Mishra, 2009; p. 63)

However, as the suggestion of several works, technology should be consistent with content knowledge (Mishra and Koehler, 2005, 2006, 2009) and the over focus on this led to the fall in pedagogy (Alvarez et al., 2009). In addition, as a heart of knowledge, TPACK has been depicted as situated, complex, multifaceted, integrative and/or transformative (Angeli & Valanides, 2009; Harris et al., 2009; Koehler & Mishra, 2009; Manfra & Hammond, 2008). To make clearer, definition and examples of TPACK were summarized (Chai et al., 2013). The authors showed the greatest concern of scholars to TPACK framework with more than 80 journal articles related up to the study time, which covered fields as theory, application, and data driven research. Depended on the specific context, solution should be flexible to combine factors interrelated (Koehler & Mishra, 2009) and it also determined by the availability of technological solutions, the familiarity of learners with the software and the instructors' pedagogical reasoning (Chai et al., 2013). To design an efficient learning activities according to TPACK, many studies started from PCK, then continued with other factors. They tried to find out the difficulties of students in learning with the subject matter as PCK, then choose the relevant technology (TPK) to help them to solve problems (Harris et al., 2010; Akkoç, 2011). Nevertheless, there are four interdependent contextual factors considered to affect TPACK designed lecture as the epistemological and pedagogical beliefs of teacher (Tsai, 2007; Kereluik et al., 2011), the interpersonal dimension of students in collaborative works (Koehler et al., 2007), cultural/institutional factors on technology (Almas & Krumsvik, 2008) and the physical/technological provision in schools (Polly et al., 2010).

To map TPACK to students, Chai et al suggested TLCK framework (Figure 2B). With this, the student's notation of particular content should be investigated first (LCK corresponding to PCK), then his/her learning with technology knowledge (TLK corresponding to TPK). Technological learning, knowledge of students influences on the teacher's decision of applied technology. TLCK (Technological Learning Content Knowledge, corresponding to TPACK) is knowledge of student how to use the proper technology in learning specific content efficiently. It's sophisticated as TPACK. They supposed that the design of lectures would be better with an understanding of student's awareness of these areas (Chai et al., 2013). Figure 2 below shows the revised presentation of TPACK to TLCK for learners.

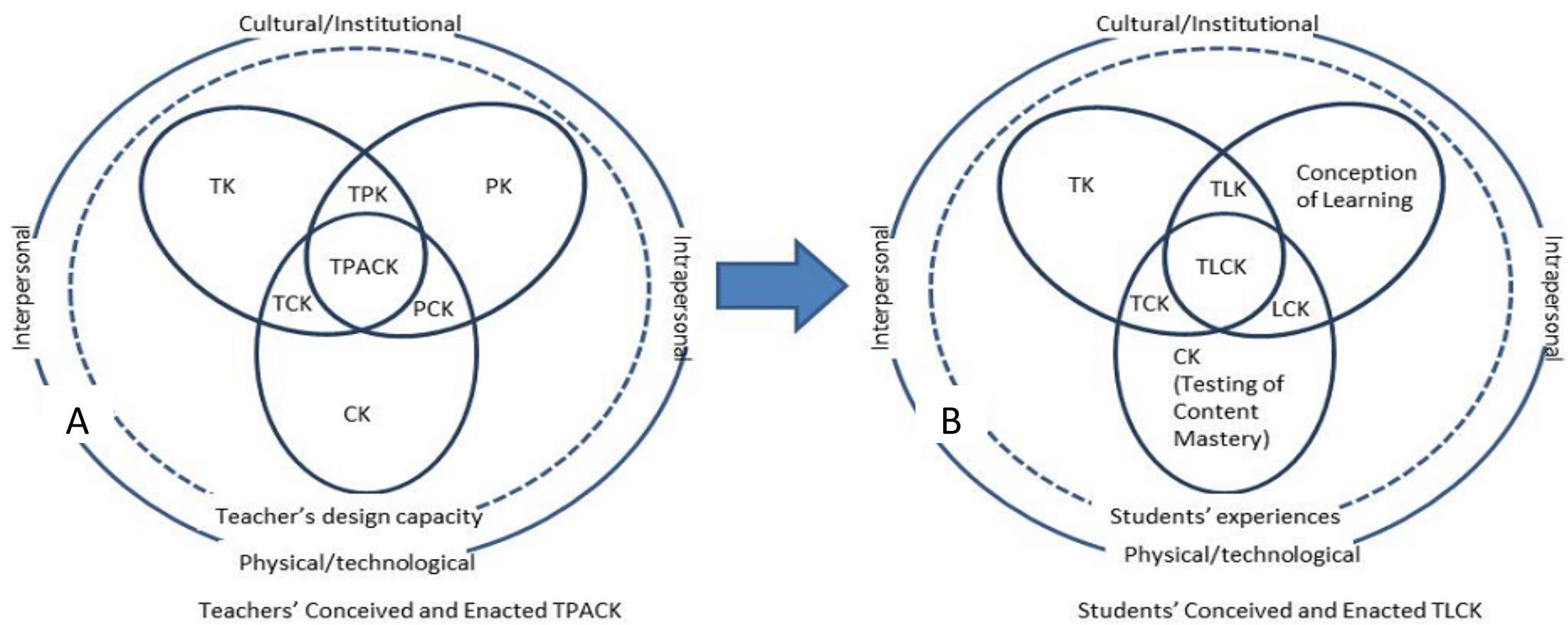

Figure 2. The revised TPACK with TLCK framework (Figure 3. In Chai et.al, 2013; p. 45)

To evaluate how successful ICT brings about to student's learning, Howland et al. (2012) proposed five dimension as: (a) active-where students were not passive listeners but actively manipulating objects and information; and observing results (b) constructive—where students constructed knowledge, reflected, and articulated their personal understandings of phenomenon (c) authentic–where students engaged in the solving of real-world problems (d) intentional—where students set their learning goals and planned their learning pathways; and (e) cooperative—where students worked with peers to learn.

*2.2 TPACK & TLCK Application Procedure*

To summarize TPACKs' components applied in the learning process, the author suggests the diagram in Figure 3. With the sequential processes, the proper knowledge required to get the success. The first phase is competency target analysis where CK, TK prior meanwhile the CK, TK and TCK in addition for the next phase: context analysis. In this phase, the teacher must analyse the learning condition as culture, current competency of the learner, current infrastructure, etc., so that the knowledge of content, technology, and technological content knowledge essential needed for the teacher in this phase. In the next three phases: learning design, learning application, learning evaluation and feedback, all mentioned knowledge should be equipped for the tutor to make successful. More than that, TPACK is required for efficiency of learning. The experience of the processes and the result of learners as well the feedback from stakeholders would enrich the kinds of knowledge of the educator.

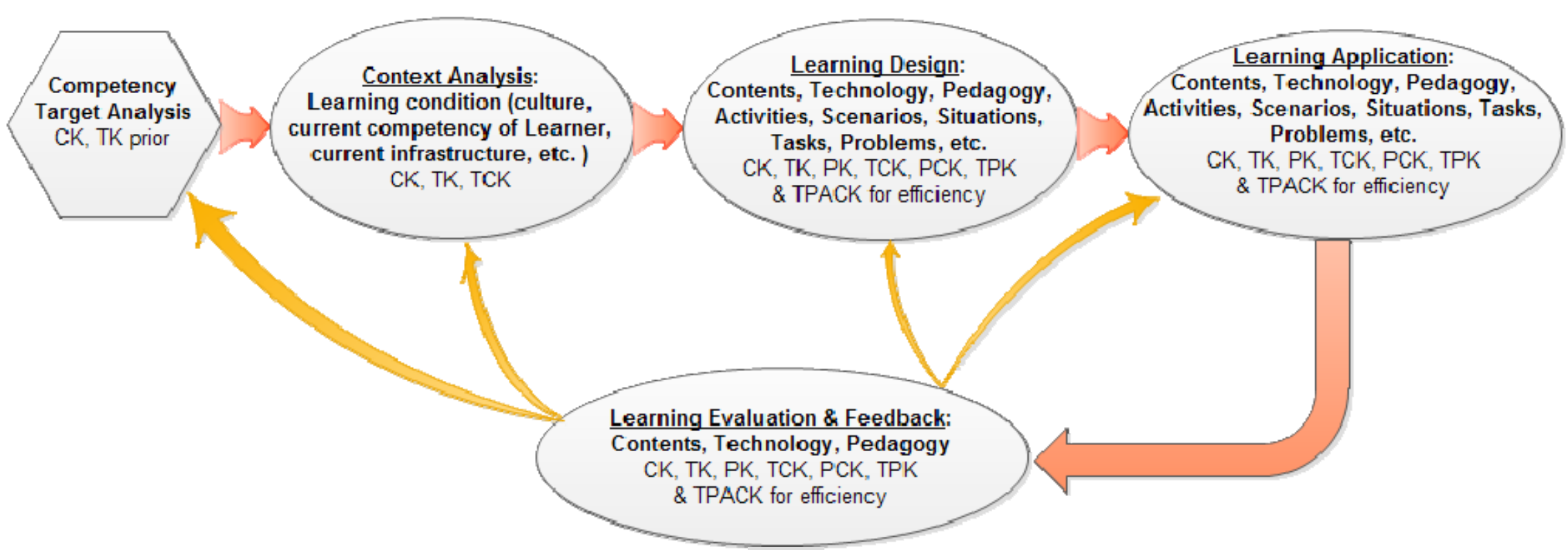

Figure 3. TPACK's components applied in the learning process for the educator

To map the qualification of learner to TLCK framework (Figure 2B), the Figure 4 is recommended. In this graph, the learner is active to follow the procedure. He/she analyses the problem target in the 1st phase where knowledge of content and technology is prior essential. His/her activities as the 1st dimension in Howland's suggestion (Howland et al., 2012). In the next phase, if the learner / student has right knowledge as CK, TK and TCK, he/she can analyse the context, in preparation for the tasks given by the instructor such as negotiation in pair/group's work, etc. In the three next stage, if the learner has the true qualification of CK, TK, LK, TCK, LCK, TLK, he/she has the advantage of doing the task well, and TLCK for the efficiency of assessments, and so on. For the student, the item "methodology" was used instead of "pedagogy" which need interact between the instructor and the student. By choosing the proper methodology can collaborate with stakeholders efficiently to accomplish the work with knowledge construction, and so forth.

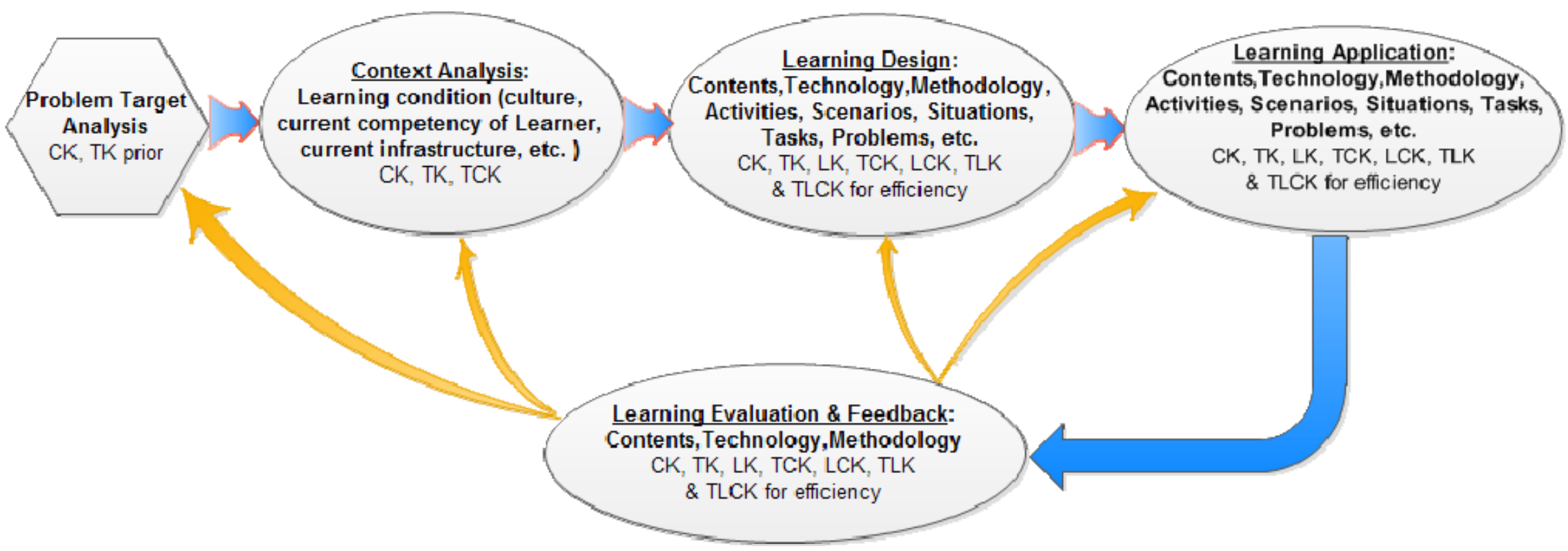

Figure 4. TLCK's components applied in the learning process for the learner

Finally, by the experience, the students would enhance the skills and competency as well.

*2.3 Active and Collaborative Learning Model (ACLM)*

In this part, the author uses the 3rd phase in the Figure 3 to choose an appropriate model. Active learning based on constructivism theory with two strands: cognitive constructivism and social constructivism. Several constructivists ideas have been used to inform adult education, whereas pedagogy applies to the education of children. There are different in emphasis, but they also share many common perspectives about teaching and learning (Jonassen, 1994). With the support of ICT, the collaboration of learner is essential (Downes, 2012; Siemens, 2004). Different from the traditional model in which listening and writing passively, students are encouraged to join in activities with the active learning model (Bonwell & Eison, 1991). They actively study learning materials such as documents, textbooks... or searching necessary contents with the help of instructor to get knowledge and skills (Newman et al., 2003). Students can work independently, in pair or group and be encouraged to present and defend result in group or class. Many researchers agree that most of learners feel excited and have advantage result due to that work (Peter et al., 2009; Godfrey, 2013; Lilia & Sonia, 2014).

In the learning model, students play a major role to find knowledge. They practice skills to solve problems, working in a team, presenting, asking, answering… Students also actively join in evaluating process included self-evaluation, group evaluation and external evaluation. By the way, they absorb knowledge, correct wrong information and learn more experiences from learning partner (classmate…) and teacher (instructor).

As an instructor, the teacher also plays an important role because giving suggestion to help student solving problem efficiently. The instructor also analyses, evaluates and corrects the wrong in cognitive process and inexactly self-evaluating, peer-evaluating of students.

The efficiency of learning activities improved with the support of ICT as various rich multimedia contents, tools for searching, building contents, evaluating, discussing etc. via the internet and social network (Lilia & Sonia, 2014; Newman et al., 2003; Downes, 2012; Siemens, 2004).

## 3. ACLM for the Introductory Physics Part 2

### *3.1 Context of the Study*

According to credit training tree, the Introductory Physics is a common and a prerequisite subject like Mathematics, Introductory Informatics for the 1st year student before registering the next business (HNUE, 2012). The Introductory Physics combines two parts: the part 1 students studied in the 1st semester, and the part 2 in the 2nd semester. Most of introductory physics contents mentioned in school with easier requirement and experience with practicing in the progress training for university entrance of leaners make them subjective feel. Many students come from rural and mountainous region therefore the skills are limited. However, they have time to learn several techniques in the Introductory Informatics course. Some of them got experiences in the Introductory Physics part 1 by using HNUE LMS without obligation. More than practice of using an LMS, the result reveals a prospective trend (Nam, 2014).

The student can adapt quickly to changing and eager to learn a new thing which exposures a suggestion to exploit this feature of learner to enhance learning effectiveness.

For that reason, a learning environment was created for the 1st year student of the Faculty of Technology Education in which the ALCM implemented with ICT supported. An online survey by using the Google form was done to get students' responses in the middle of the course.

By analysing context as the 2nd phase in the Figure 3 above, the author has exploited the HNUE LMS in training which build on Moodle with ALCM.

*3.2 Deploying ACLM with ICT Support*

A/ Management of learning activities: In the work of Hazzan et al. 2011, the authors suggest 4 phases in ACLM process: preparation → learning activities → discussion → summary. The details of activity depend on the learning condition and learning content. The author goes further in the Introductory Physics Part 1, then adjust in the Part 2 with the help of TPACK and TLCK analysis as follows:

Phase 1: preparation. Due to the problematic requirement, student studies, individually or in small group. The instructor announces tasks, guides the way to study. The instructor has CK, TK of using an LMS, exploiting internet & ICT, group working etc. to guide. The guided materials, assessment… prepared in various types as documents, rich media elements as a video-clip, pictures… and so on, so that TCK is required. PCK is needed to select enough data for the task, for example. The instructor also has PK of ACLM to use this model. TPK is asked to use tools for supporting methodology in the model as problem-based inquiry… TPACK is required to combine method for the model efficiently.

As analysing the TLCK in Figure 4 above, CK & TK needed for students to understand the problem as a task given by the instructor. These knowledges support student to judge the target problem.

Students should have CK, TK, TCK to analyse learning condition. If working in groups, students negotiate each duty.. They can do in class for a live assessment or via the internet by using separate online forum of HNUE LMS or Facebook etc. for a home assessment. The number of members in a learning group should be chosen as 3-5 due to a reasonable duty arrangement, in our point of view. Working process must be presented by the instructor.

Activities of the instructor and the student mapped to phase 1st and 2nd according to the Figure 3 & 4.

Phase 2: learning activities. The student is actively studying given problem, searches documents, etc. He/she works individually to solve own problem. The activities correspond to the phase 3rd in Figure 3 & 4. Preparing materials and designing own obligation work, students should have CK, TK, LK, TCK, LCK, TLK to accomplish their work. If they have TLCK, the work would be done efficiently. Of course, as in other phases, such knowledges have not been created immediately but by the experience: doing & learning.

As an instructor, the teacher does not interfere in details of learning design. However, he/she estimates the procedure and give instruction for students to do. Depending on the context of study, there are several applied methods as IMS learning design and pattern languages (Lockyer et al., 2008). Note that kinds of necessary knowledges for the teacher was studied above.

Phase 3: discussion. Students discuss in pair or own group to make an addition, correct the project outcome. They do work online or offline. Writing down questions about own concerned subject need to be answered by other groups and corrected by the instructor.

This phase maps to phase 4th in Figure 3 & 4. All activities done in the learning space. Required knowledges are the same analytic ones in the phase 3.

Phase 4: summary. Students present and defend the result by individual or group. Encourage students in a group presenting in combination. In this mode, each student takes a role as speaker for a part of speech in sequence. All groups debate in class to find out solution for all relevant subjects.

Finally, the instructor summarizes, gives added information, answers question, corrects the results of individual, duet or groups.

In fact, this phase, according to Hazzan et al., combines both phase $4^{th}$ & $5^{th}$ in Figure 3 & 4. Due to the activities, all knowledge requirements are needed and for the efficiency in learning, TPACK for the instructor, TLCK for the student in consequence as well.

B/ Learning content design: Method of pattern languages is taken to redesign learning content which based on formal documents as textbooks, handbooks, etc. or learned from informal resources as searching from internet (Hung & Nam, 2013). Classification and purpose uses of the pattern languages method were presented in another work of the author (Nam & Quyen, 2014). In the introductory physics course, temporary sample is applied due to the consistency with constructivism. The student is required to rebuild content for presentation according to the structure as follows (modified and extended structure of the author's other research (Nam, 2014):

1/ Group Name

2/ Member Name

3/ Name of Subject

4/ Comic: Graph related to the subject

5/ Context: A sample related to the field

6/ Target and Purpose: Target and purpose using a subject

7/ Features and applying condition: features and applying a condition of the subject

8/ Applying: activities, problems (exercise) solving, samples of subject application in life, technology, etc.

9/ Extra information: Addition about subject or suggestion finding more information with related subjects

10/ Referrences: resource referenced for content preparation

11/ Self-evaluation (in group mode): student's task in group and evaluated weight mark by own group due to the contribution and effectiveness.

In the view of TCLK framework, knowledge of students developed as follows: To construct content of this structure, CK is required for the student to understand the related content at first. He/she has TK to see what technology to do the work, as using software, internet, etc. Other knowledges as LK, TCK, LCK, TLK needed to study with the help of instruction given by the instructor. LK is the conception of learning according to the Figure 2B, depicted as the procedure to rebuild the learning content, for example. TCK is knowledge of using technology related to build learning content as drawing, creating a presentation, etc. LCK is knowledge of rebuilding the learning content according to constructivism. TLK is knowledge of technology to study methodologically, as using calculator to convert units for the exercise solving method. And the last, TLCK is the knowledge to use technology to construct the learning content efficiently as using simulation, rich media data in a combination of the content reasonably, for instance.

C/ Learning evaluation: formative and summative evaluation both used. At the conclusion of the course, students must participate in the final exam according to the training required to obtain a final score. A portion of an exam of theory question follows the above structure (without the item $10^{th}$ and $11^{th}$).

A middle mark of the student got by weighted average mark of individual and group activities. Activities include solving problems, making project, doing exercise, doing a middle exam, etc. which announced in class.

Evaluation points of individual and group are published. Members of a group negotiate evaluating weight point due to each contribution. A students mark for a group solved problem is equal to the weight point multiplied with groups mark, according to the framework of rubric evaluation (Goodrich, 1996; Laura, 2010). The table 1 below shows rubric evaluation for group assessment:

Table 1. Rubric evaluation items for group assessment

| | Items | Point | | Items | Point |
|---|---|---|---|---|---|
| Software package (as assessment) | Idea | 10 | Activities | Discussion in online group | 10 |
| | Exact | 10 | | Discussion in live class | 15 |
| | Presentation | 10 | | Question | 10 |
| | Application | 20 | | Oral presentation and defence | 15 |

As shown above, all types of knowledges mentioned in the Figure 4 are needed for the students to finish their work. They should know that to limit the lost points due to misunderstanding or lack of related knowledge, as using the LMS for online discussion, etc., for examples.

D/ ICT support: The efficiently ICT support is clearly shown in building rich multimedia content, searching, collaborated process, presenting, etc. The author has deployed the Learning Management System (LMS) of HNUE to make the online introductory physics course to support a student at the address: (http://lms.hnue.edu.vn) with the mode of 'Blended eLearning' (Friesen, 2012).

The course contains required subjects for individual or groups as task or assessment. The student is required to add a submission via LMS. A presentation called software package was in the form of Word or PowerPoint document due to the rich content. All of requirements, evaluation guide, group working guide, etc. were published. There are two kinds of forum was created for each subject as an assessment. The first is a common forum for whole class discussion, the second is separate for group's private work. Beside talking about physics contents, the instructor and members can share documents, etc. By viewing the separate forum, the instructor can value member's works. Extra information also was uploaded as attached documents or from other resources as videos for physics experiments, technology and website links related and so forth. The LMS system supports conference room for online discussing between students and instructors at a timely decision.

There are changes in learning organization between the part 1 and part 2 of the Introductory Physics course. The student was encouraged to use LMS in the part 1, whereas taking part in activities via LMS is an obligation. The assessment could be a hard copy or photo uploaded via LMS in the part 1, but a software in the part 2. It makes a variety of rubric evaluation items as above. The basis of change is the lack of ICT skills of students in the Introductory Physics part 1. However, their experiences in using LMS in the part 1 of Introductory Physics and Introductory Informatics course accounts for the modification.

For efficient learning, instructor guides, student exploiting other ICT services as sharing documents, building contents, etc. via the internet with the help of Google Drive, Drop box… and social networks as Facebook to get information and support quickly and in time because as the survey result shown students pay a great attention to communicate with social networks. The knowledges of TK, TCK, TLK & TLCK mentioned in the Figure 4 are needed for the students to exploit ICT to finish their work.

### *3.3 Several Problems in Deploying ALCM and Discussion*

A/ Deploying result: the author has deployed ALCM for the introductory physics part 2 and surveyed students at the middle of the course by using the Google form. The survey was carried out via the Internet to keep private and separate feedbacks. With 82 responded votes, the result as follows:

The student was asked to self-evaluate his/her competency in several areas and the betterment of learning from the beginning to the current time of the introductory physics (IPC) part 2. There are five options in the Likert scale with weight: Totally Disagree (1), Disagree (2), Neutral (3), Agree (4), Totally Agree (5). The sample statistics were calculated as Equation 1 as follows (http://en.wikipedia.org/wiki/Weighted_arithmetic_mean):

*Equation 1. Sample Statitics calculated*

Sample weighted mean:

$$\overline{x} = \frac{\sum_{i=1}^{n} \omega_i x_i}{\sum_{i=1}^{n} \omega_i} \quad (a)$$

Sample weighted standard deviation:

$$s_d = \sqrt{\frac{\sum_{i=1}^{n} \omega_i (x_i - \overline{x})^2}{\sum_{i=1}^{n} \omega_i - 1}} \quad (b)$$

Sample weighted standard error:

$$s_e = \sqrt{\frac{\sum_{i=1}^{n} \omega_i (x_i - \overline{x})^2}{\sum_{i=1}^{n} \omega_i \left( \sum_{i=1}^{n} \omega_i - 1 \right)}} \quad (c)$$

Where $\omega_i$ is the weight for the ith observation, $x_i$ is the value of the ith observation, n is the number of observations.

Here are the statistical results:

Table 2. Student self-evaluated competency before joining IPC Part.2 (note: the rate was rounded up to 2 digits in decimal place)

| Competency/Skills | Sample weighted mean | Sample weighted standard deviation | Sample weighted standard error |
|---|---|---|---|
| 1. Proficient in Word | 3.67 | 0.92 | 0.09 |
| 2. Proficient in PowerPoint | 3.16 | 1.18 | 0.15 |
| 3. Proficient in exploiting the Internet | 3.89 | 0.89 | 0.09 |
| 4. Proficient in Facebook for learning purpose | 3.61 | 0.99 | 0.11 |
| 5. Proficient in HNUE LMS | 3.28 | 1.16 | 0.15 |
| 6. Proficient in analysing, evaluating, discussing of physics contents | 3.00 | 0.96 | 0.10 |
| 7. Proficient in solving physics excercise | 2.72 | 0.93 | 0.10 |
| 8. Proficient in independent working | 3.55 | 0.92 | 0.09 |
| 9. Proficient in collaboration | 3.60 | 0.89 | 0.09 |
| 10. Proficient in organising collaboration | 3.49 | 0.86 | 0.08 |
| 11. Proficient in presentation | 3.13 | 1.09 | 0.13 |
| 12. Proficient in defending | 2.90 | 1.00 | 0.11 |
| 13. Proficient in questioning | 2.87 | 1.07 | 0.13 |
| 14. Proficient in criticising | 2.94 | 1.03 | 0.12 |

The questions mentioned types of knowledge of students as TK (Q1-Q5), CK (Q6, Q7), TCK (Q1, Q2, Q6, Q7), TLK (Q4, Q5), LK (Q8-Q10), TLK & TLCK (Q6-Q14).

The sample weighted mean of table 2 can be divided into 3 groups: negative ($m<3$), neutral ($3 \le m < 3.4$), and somewhat positive ($3.4 \le m$). The weak skills students evaluated are solving physics exercise, defending, questioning and criticising. That caused by the domination of negative rate (totally disagree and disagree) / positive (agree and totally agree) rate of each student's skill at 41.4/19.5 (solving exercise), 31.8/28.1 (defending), 39.1/29.3 (questioning) and 36.6/32.9 (criticising). The neutral votes are skills of using PowerPoint and LMS, analysing, evaluating, discussing of physics contents, collaboration and presentation. The rates between negative and positive votes are: 29.2/37.9 (PowerPoint), 20.8/42.7 (HNUE LMS), 30.5/26.8 (analysing…), 24.4/39 (presentation). The somewhat positive results are skills in Word use, exploiting the Internet, Facebook use, collaboration and organising collaboration. The domination of positive votes over negative ideas made this: 61/8.5 (Word), 70.7/6.1 (Internet), 58.5/11 (collaboration), 51.2/13.4 (organising collaboration). The sample weighted standard deviation measures the amount of variation or dispersion of the average. A low standard deviation indicates that the data points tend to be very close to the mean, while as a high standard deviation indicates that the data points are spread out over a large range of values. The result reveals the split between students in evaluation.

The reason accounts for the fact that students got skills by being trained in the Introductory Informatics course, whereas they learnt Word, Excel, Internet (not PowerPoint). They reached other achievements from the IPC Part.1 (Nam, 2014) and somewhat from the environment of learning and entertainment as well. The high neutral index in comparison with others shows the requirement of improving learning facilities.

At the time of the survey, students thought their skills as follow:

Table 3. Student self-evaluated competency at current time of IPC Part.2 (note: the rate was rounded up to 2 digits in decimal place)

| Competency/Skills | Sample weighted mean | Sample weighted standard deviation | Sample weighted standard error |
|---|---|---|---|
| 1. Improved using Word skill | 3.99 | 0.75 | 0.06 |
| 2. Improved using PowerPoint skill | 3.71 | 0.87 | 0.08 |
| 3. Improved using an Internet skill | 3.96 | 0.71 | 0.06 |
| 4. Improved using Facebook for learning skill | 3.96 | 0.69 | 0.05 |
| 5. Improved using the LMS HNUE skill | 3.94 | 0.69 | 0.05 |
| 6. Improved skill of analysing, evaluating, discussing physics contents | 3.55 | 0.88 | 0.08 |
| 7. Improved solving physics exercise skill | 3.15 | 0.94 | 0.10 |
| 8. Improved independent working skill | 3.87 | 0.83 | 0.08 |
| 9. Improved collaborating skill | 3.74 | 0.84 | 0.08 |
| 10. Improved organising collaboration skill | 3.68 | 0.78 | 0.07 |
| 11. Improved presenting skill | 3.67 | 0.88 | 0.08 |
| 12. Improved defending skill | 3.35 | 0.92 | 0.09 |
| 13. Improved questioning skill | 3.41 | 0.92 | 0.09 |
| 14. Improved criticising skill | 3.38 | 0.91 | 0.09 |

The outcome of table 3 depicts the improvement of a student's skill. All samples weighted means are greater than 3, especially several skills improve clearly with the value of weighted mean next to 4. The result (plus/negative idea): 82.9/3.6 (using Word), 69.5/7.4 (using PowerPoint), 84.2/3.6 (using Internet), 79.3/2.4 (using Facebook), 78.1/2.4 (using HNUE LMS), 79.3/4.9 (independent working), 74.4/11 (collaborating), 69.5/8.5 (organising collaboration), 69.6/7.3 (presenting), 45.5/11 (defending), 48.7/11 (questioning), 47.5/12.2 (criticising). Though the number of neutral ideas is equal to the agreement for several polls, the author can see the clear enhancement of the student's soft-skills due to the experience in learning activities. The competency of solving physics exercises also increases, but not as high as others. Therefore, it will be more concerned in the post about the course.

The student was also asked to judge the documents, contents, methodology, instructor's responsibility etc. of the course:

Table 4. The student evaluates on HNUE LMS and current methodology at current time of IPC Part.2 (note: the rate was rounded up to 2 digits in decimal place)

| Items | Sample weighted mean | Sample weighted standard deviation | Sample weighted standard error |
|---|---|---|---|
| 1. Documents meet student's requirement | 3.41 | 0.87 | 0.08 |
| 2. Topics meet student's capability | 3.46 | 0.97 | 0.10 |
| 3. Analysing topic into structure make the content clear | 3.76 | 0.90 | 0.09 |
| 4. Analysing application of topic makes content reality | 3.84 | 0.78 | 0.07 |
| 5. Analysing technical application of topical supports specific subjects better | 3.74 | 0.84 | 0.08 |
| 6. Solving exercise in presenting a topic improves solving skill | 3.51 | 0.88 | 0.09 |
| 7. Preparing duration (1 week) for topic suitable | 3.49 | 1.06 | 0.12 |
| 8. Common forum suitable | 3.71 | 0.76 | 0.06 |
| 9. Separate group forum helps group working proficient | 3.46 | 0.85 | 0.08 |
| 10. Public evaluation, weekly suitable | 4.35 | 0.57 | 0.04 |
| 11. Instructor's feedback in time | 4.01 | 0.66 | 0.05 |
| 12. Instructor's answers, meet student's requirement | 3.85 | 0.83 | 0.08 |
| 13. Assessments deliver & feedback via HNUE LMS suitable | 3.68 | 0.93 | 0.10 |
| 14. Organising group framework suitable | 3.66 | 0.89 | 0.09 |
| 15. Organising group framework evaluates member exactly | 3.45 | 0.94 | 0.10 |
| 16. Independent working with topic suitable | 3.62 | 0.94 | 0.10 |
| 17. Learning framework evaluates student's competency, exactly | 3.49 | 0.79 | 0.07 |
| 18. Methodology makes student interested | 3.51 | 0.85 | 0.08 |
| 19. Methodology enhanced student's activeness | 3.76 | 0.87 | 0.08 |
| 20. Methodology enhanced self-study competency of students | 3.88 | 0.81 | 0.07 |
| 21. Methodology helps student learning proficiently | 3.40 | 0.89 | 0.09 |
| 22. HNUE LMS friendly and easy use | 3.62 | 0.91 | 0.09 |

As the result shown, students still kept optimistic attitude to learning environment and methodology with the sample weighted means are above 3, whereas many value items in somewhat positive and positive. The outcomes are (positive/negative): 52.4/14.6 (document satisfied), 64.3/18.3 (methodology satisfied), 73.1/8.6 (analysing method agreement), 78.6/6.1 (analysing application agreement), 67.1/7.3 (analysing technical application agreement), 53.7/12.2 (solving exercise agreement), 63.4/19.5 (preparation duration time agreement), 71.9/6.1 (common forum suitable), 52.4/10.9 (separate online forum agreement), 95.1/0 (public evaluation agreement), 84.1/2.4 (feedback in-time satisfied), 74.4/4.8 (feedback satisfied), 64.7/8.6 (framework of assessment and feedback agreement), 64.6/9.7 (organising group agreement), 53.7/18.3 (group evaluation agreement), 45.1/4.8 (methodology interested), 68.3/8.5 (enhanced students activeness), 76.8/6.1 (enhanced self-study), 45.2/12.2 (enhanced learning effectiveness), 62.2/10.9 (LMS friendly and easy used).

The neutral index rate of several items of two table 3 and 4 was highly in contrast with others. More or less of them were related to soft-skills as defending (40.2 & 43.9), questioning (31.7 & 40.2), criticising (30.5 & 40.2), while others associated to methodology of teaching/learning and evaluating. Students, however, still feel trouble in working with physics contents, as ideas the author received from the survey for input items.

To find out what is disadvantage in learning, the author asked students extra information as follows:

Table 5. Reason for mistakes in analysing physics contents at current time of IPC Part.2 (note: the rate in percentage was rounded up to 1 digit in decimal place)

| Items | Percentage |
|---|---|
| Learning content in text book difficultly understood | 26.3 |
| Difficult topic | 19.0 |
| Not enough information | 17.3 |
| Lack of self-studying preparation | 12.8 |
| No experience | 24.6 |

Table 6. Reasons for limited discussing in the common forum at current time of IPC Part.2 (note: the rate in percentage was rounded up to 1 digit in decimal place)

| Items | Percentage |
|---|---|
| Don't know how to start | 63.0 |
| Not interested in discussion | 14.1 |
| Don't know how to implement discussion | 21.7 |
| Afraid of being evaluated by peer | 1.1 |

The facts from two tables indicate that the instructor needed to help students more to solve problems such as guiding, or training them to find information, giving suggestion, encouraging by somehow even evaluation. Several thoughts the author got pointed out the general students' mental reliance on others or instructor. Many students are not willing to do a task if it was not obligated. Moreover, the difficulties come from other causes as economic conditions, psychology, ability etc. of students.

The author asked students about their devices to prepare learning content and access to the HNUE LMS:

Table 7. Devices for learning content prepared at current time of IPC Part.2 (note: the rate in percentage was rounded up to 1 digit in decimal place)

| Items | Percentage |
|---|---|
| Personal computer | 50.0 |
| Borrowed computer | 26.5 |
| Rent computer | 7.8 |
| Mobile devices | 15.7 |

Table 8. Devices for accessing HNUE LMS at current time of IPC Part.2 (note: the rate in percentage was rounded up to 1 digit in decimal place)

| Items | Percentage |
|---|---|
| Personal computer | 51.5 |
| Borrowed computer | 26.7 |
| Rent computer | 6.9 |
| Mobile devices | 14.9 |

The number shown that 1/3 1st year students have not owned computer or device to prepare learning content or joining LMS. Because HNUE's students are educated to become a teacher, so that they are sponsored by the government which leads many come from rural and mountainous region. Some feedbacks in open forms revealed the difficulties of sharing a computer with other members with family or accessing due to the limit of Internet bandwidth or weakness of wifi etc., so that 'Blended-learning model' should be better chosen than e-learning model.

The author also questioned students about studying duration for a week at university:

Table 9. Students learning days at university at current time of IPC Part.2 (note: the rate in percentage was rounded up to 1 digit in decimal place)

| Items | Percentage |
|---|---|
| 1 day | 0 |
| 2 days | 0 |
| 3 days | 10 |
| 4 days | 37.5 |
| 5 days | 41.3 |
| 6 days | 11.3 |

A half of surveyed students take part in different credit courses at university (52.6% for 5 and 6 studying days) whilst others have time out of school. They revealed their activities out of school time as follows:

Table 10. Student activities out of learning time at university at current time of IPC Part.2 (note: the rate in percentage was rounded up to 1 digit in decimal place)

| Items | Percentage |
|---|---|
| Home self-study | 25.8 |
| Library self-study | 11.2 |
| Collaboration study | 13.1 |
| Sport activities | 12.4 |
| Entertainment activities | 19.1 |
| Social activities | 5.2 |
| Earning works | 13.1 |

Table 11. Students' major activity outside school at current time of IPC Part.2 (note: the rate in percentage was rounded up to 1 digit in decimal place)

| Items | Percentage |
|---|---|
| Home self-study | 58.5 |
| Library self-study | 2.4 |
| Collaboration study | 2.4 |
| Sport activities | 6.1 |
| Entertainment activities | 15.9 |
| Social activities | 2.4 |
| Earning works | 12.2 |

The consideration of students various in many fields, most of them reserved for studying (50.1% in table 9 and 63.3% in table 10). Others pay attention to different activities in combination or separately. This may explain for the difficulties of students in studying with ALCM because the lack of time preparing for a topic, for example. Some of them faced with the trouble in working, negotiation for the group framework, as well.

The author measured the frequency ICT use of the student. It shows us:

Table 12. Students frequency use of software at current time of IPC Part.2 (note: the rate in percentage was rounded up to 1 digit in decimal place)

| Frequent use of softwares | Word | PowerPoint | Excel | Facebook | HNUE LMS | Other software |
|---|---|---|---|---|---|---|
| None | 0 | 8.5 | 9.8 | 1.2 | 0 | 8.5 |
| Rarely | 7.3 | 25.6 | 36.6 | 1.2 | 6.1 | 22 |
| Sometimes | 47.6 | 42.7 | 46.3 | 15.9 | 42.7 | 41.5 |
| Frequently | 32.9 | 17.1 | 4.9 | 46.3 | 42.7 | 22.0 |
| Always | 12.2 | 6.1 | 2.4 | 35.4 | 8.5 | 6.1 |

Consequently, the active rate of using ICT for tasks increased, as in table 12. Because Excel is not essential for the presentation, so that the fact of uses is less than other software. Though using HNUE LMS was forced for duty which leads to the high frequency uses, the positive attitude of students to Facebook is especially impressive. Facebook is not only applied for entertainment as chit-chat, but also exerted for work because of friendly and quick response. In addition, some software was exploited for studying the content as simulation, design or draw, etc. Open question created to find out the reasons for limiting discussion on LMS and several students gave their thought as:

*"I don't know how to start a discussion. I'm afraid of being ashamed by criticism"*—Student No. 16.

*"I use Facebook frequently and it's easier for me to use Facebook to discuss with classmate and friend"*—Student No. 23.

*"I don't have a PC so that it's difficult for me to borrow. I have to hire PC in the Internet café to do my work so that it's a trouble to follow the discussion"*—Student No. 38.

*"Sometimes it takes so long to access LMS. The WiFi signal in the hostel is weak. It doesn't announce the news as Facebook for discussion"*—Student No. 51…

Art performance of students (poetry, music…) in the rest time or at the welcome session to reduce stress and enhance learning interest is one of the solutions mentioned in the author's another study (Nam, 2013; 2014). This was supported by almost students.

Thus the reason for students' interest in ACLM was confirmed by the active learning and developmental skills with activities which is essential for the future job. Some students have not adapted with new learning methodology and relied on others. Some of them felt ashamed to make a presentation. These results came from other open questions. The disadvantage of psychology should overcome with practice.

B/ Suggestion: ACLM requires the timely support of instructor to students. The student should be trained to have skills of self-solving problem and collaboration in learning. To meet such requirements, the facilities of ICT play important roles.

The LMS of HNUE is based on a well known open-source named Moodle, which supplies free services for courseware. It is necessary to research and invest to improve friendly facilities of LMS as registration, using resources for both student and instructor. According to the students' responses an the author's experience of exploting LMS, if an application form connected with social networking as Facebook studied, the information of closure would be transferred to the student and instructor timely which enhance the efficiency of learning. The integration of Moodle & Facebook in education is also the suggestion of the other works (Katherin, 2012; Natasa et al., 2014), especially there has a Facebook package for the Moodle (http://docs.moodle.org/20/en/Facebook_package). The university should support more utilities for students as installation more wifi devices, employing computer-lab and reservation bandwidth for accessing education resources as LMS.

**4. Conclusion**

Applying ACLM in the introductory physics course shows the advantage in competency-based education. Using ACLM with the TPACK framework in specific subjects training makes efficient in learning because students get benefit from active and collaborative working in which many soft skills for future job development. Deploying ICT in combination of social network and LMS with blended-learning mode will increase the efficiency of ACLM so it need to be studied and supported.

**Acknowledgment**

The author thanks to the support of the project SPHN-13-314.